\documentclass[msmath,amssymb,aps,prapplied,twocolumn,epsfig,showpacs,bibliography,lengthcheck,superscriptaddress]{revtex4-1}

\usepackage{graphicx}
\usepackage{dcolumn}
\usepackage{bm}
\usepackage{hyperref}
\usepackage{epstopdf}
\usepackage{subcaption}
\usepackage{caption}
\usepackage{footmisc}
\usepackage{amsmath}
\usepackage{color}

\begin{document}
\title{Light-shift-free and dead-zone-free atomic orientation based scalar magnetometry using a single amplitude-modulated beam}%
\author{Q.-Q. Yu}
\author{S.-Q. Liu}
\author{C.-Q. Yuan}
\author{D. Sheng}
\email{dsheng@ustc.edu.cn}
\affiliation{Department of Precision Machinery and Precision Instrumentation, Key Laboratory of Precision Scientific Instrumentation of Anhui Higher Education Institutes, University of Science and Technology of China, Hefei 230027, China}

\begin{abstract}
Detection dead zones and heading errors induced by light shifts are two important problems in optically pumped scalar magnetometry. We introduce an atomic orientation based single-beam magnetometry scheme to simultaneously solve these problems, using a polarization-reversing and path-bending Herriott cavity. Here, a reflection mirror is inserted into the cavity to bend the optical paths in the middle, and divide them into two separated orthogonal regions to avoid the detection dead zone. Moreover, half-wave plates are added in the center of each optical region, so that the light polarization is flipped each time it passes the wave plates and the light shift effects are spatially averaged out. This operation is demonstrated to eliminate the unnoticed heading errors induced by ac light shifts. The methods developed in this paper are robust to use, and easy to be applied in other atomic devices.
\end{abstract}
\maketitle

\section{Introduction}
Optically pumped scalar magnetometers resolve atomic Zeeman splittings to extract the bias field magnitude ~\cite{smullin2009}. With developments in experiment schemes, hardware and techniques, scalar magnetometers have reached great advances in sensitivities~\cite{sheng2013,lucivero2021,lucivero2021b}, stabilities~\cite{wilson2019}, and miniaturization~\cite{kitching2018}. They have been widely used in researches on fundamental physics~\cite{abel2020,gnome2021}, geophysics~\cite{prouty2013}, space science~\cite{acuna2002}, and neuroscience~\cite{limes2020,zhangrui2020}. However, due to working principles, they usually suffer two important problems in practical applications, which are detection dead zones and heading errors.

As initially studied by Bloom~\cite{bloom1962}, detection dead zones in scalar magnetometers refer to certain directions of the target field where the sensor signal-to-noise ratio is significantly reduced. For example, the detection dead zones of magnetometers using dc pump beams involve regions perpendicular to the pump beam direction, and the dead zones of double-resonance magnetometers include extra regions near the radio-frequency excitation field direction. Several methods have been developed to solve this problem, including the use of pulse laser to increase the pump beam power~\cite{limes2020}, employing multiple vapor cells~\cite{acuna2002,cheron2001}, and combining multi-harmonic information~\cite{he2021}.

Heading errors correspond to the effect that the measured field magnitude is correlated with the sensor orientation.  Nonlinear Zeeman effects~\cite{bao2018,lee2021} and light shifts~\cite{oelsner2019} are two main sources of such errors, where the latter one can be the dominant factor when the bias field is not large. In optical pumping experiments, light shifts correspond to the changes of atomic ground state transition frequencies due to interactions between atoms with light beams~\cite{cohen1962,happer72}. There are two physical mechanism to generate such shifts: one is due to the transfer of coherence between ground and excited states in atomic transitions~\cite{cohen1962,norris2012}, and the other one is due to ac Stark shifts~\cite{mathur1968,happer72}. In atomic cell experiments filled with quenching gases, the former one is largely suppressed~\cite{appelt98}. The ac Stark shift can be divided into scalar, vector and tensor parts~\cite{happer72}, among which the vector part is the most important one in atomic orientation based magnetometers. This vector light shift is determined by the circular polarization part of the light, and its effect on atoms is equivalent to an effective magnetic field~\cite{mathur1968}. It is common to circumvent this problem by either locking the beam on resonance, or employing the atomic alignment based magnetometry~\cite{abel2020,zhangrui2022}.

In miniaturized atomic devices~\cite{kitching2018}, it is often preferred to leave the laser wavelength unlocked, so that both the power and space can be saved. Moreover, in certain atomic orientation based magnetometry, such as the one using a single elliptically polarized light~\cite{shah09}, the optical pumping beam has to be kept off resonance. In these cases, it requires to develop new methods and schemes to simultaneously eliminate the detection dead zones and light shift, even when the light frequency is off resonance or drifting. A previous successful scheme employs a single polarization modulating beam, which can simultaneously excite orientation and alignment resonances~\cite{benkish2010}. While the combination of these two kinds of resonances can eliminate the detection dead zones, the light shift effect is time averaged out. However, in practice, the electrical-optical modulator used for polarization modulations is sensitive to the environment parameters. This not only is a problem for the device calibration, but also can cause drifts or fluctuations of the time-averaged light shift result.

In this work, we present an atomic orientation based scalar magnetometer, which only uses a single amplitude-modulated beam. This sensor is assisted by a Herriott cavity, which is modified to lift the detection dead zones and suppress light shifts. Following this introduction, Sec. II introduces the setup and working principle of the magnetometer employing a conventional Herriott cavity, Sec. III presents the measurement results and identifies the problems in this conventional magnetometer, Sec. IV. describes the solutions of these problems by modifying the optical properties of the beams inside the cavity, and Sec. V concludes the paper.

\section{Magnetometer setup and working principle}
A Herriott-cavity-assisted vapor cell, which is made by the anodic bonding technique~\cite{cai2020,hao2021}, is used for the atomic magnetometer. This cavity consists of two cylindrical mirrors with a curvature of 100~mm, a diameter of 12.7~mm, a thickness of 2.5~mm, and a separation of 19.3~mm. In this work, the multipass cavity increases the interaction length between light and atoms~\cite{li2011} so that the cell can be kept at a relatively low working temperature, which is essential for the stability of some optical elements as discussed in the latter part of the paper. Such a cavity also provides a platform for manipulating the optical properties of beams inside.  The vapor cell, filled by Rb atoms with natural abundances and 350 torr of N$_2$ gas, is placed in a 3D printed optical platform, and heated to a temperature around 55$^\circ$C by running ac currents through ceramic heaters.

\begin{figure}[htb]
\includegraphics[width=3in]{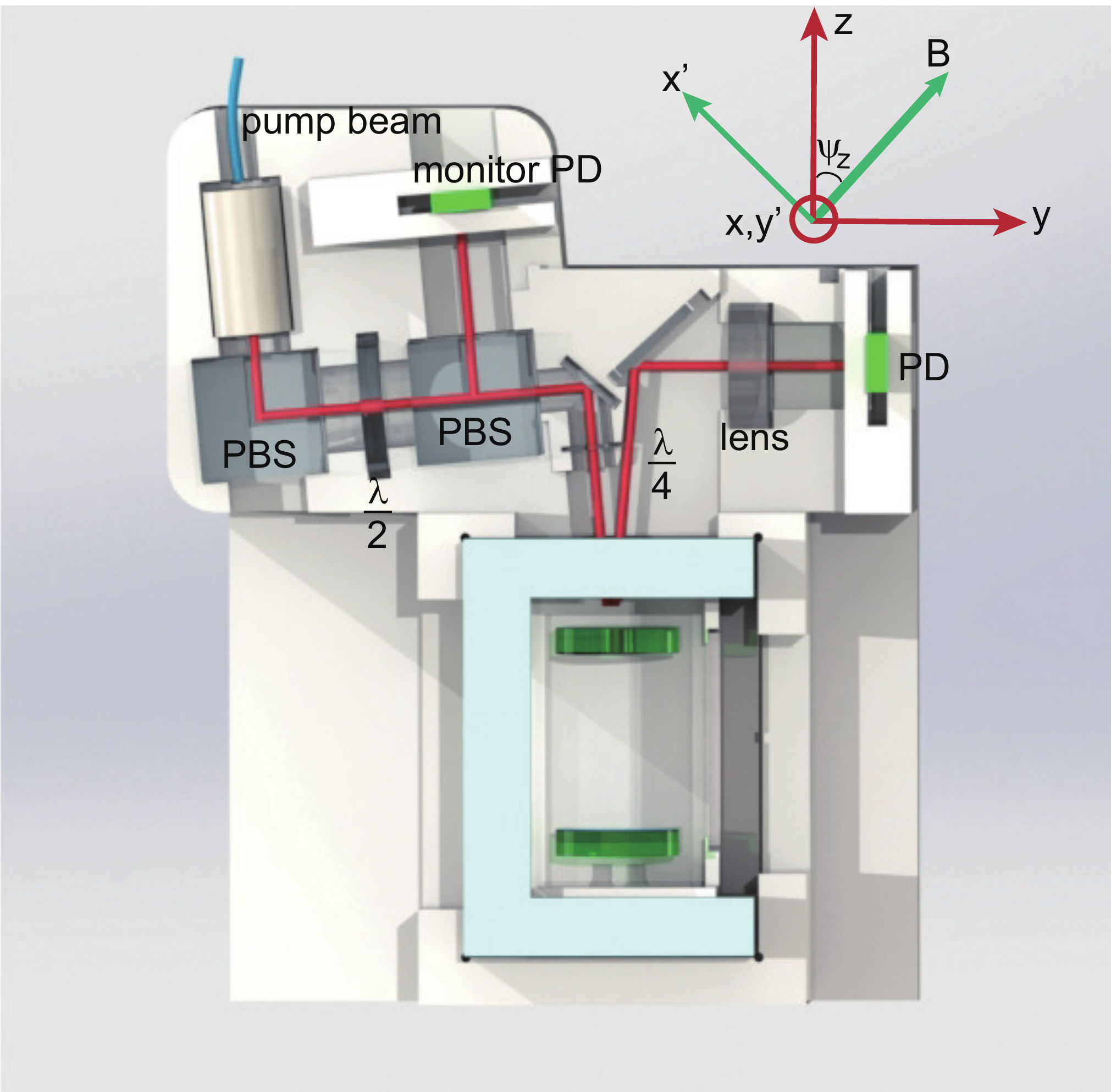}
\caption{\label{fig:scalarsetup}(Color online) Illustration of the optical platform for a single-beam Bell-Bloom scalar magnetometer. PBS: polarization beam splitter, PD: photodiode detector.}
\end{figure}

Figure~\ref{fig:scalarsetup} shows the optical platform used for a single-beam Bell-Bloom scalar magnetometer~\cite{bell1961}. The pump beam is amplitude modulated using acoustic-optical modulators with a duty cycle of 20$\%$, and fiber coupled to the optical platform. This pump beam, with a diameter of 1~mm and a circular polarization, enters the cavity from a 2.5~mm hole in the center of the front mirror, and exits from the same hole after 21 times of reflections. This magnetometer sensor sits in the middle of five-layer mu-metal shields, with a 10~$\mu$T bias field generated by solenoid coils inside the shields. The sensor is also connected to a rotation table outside the shields through a plastic holder, so that the relative angle between the pump beam direction and the bias field can be precisely controlled.

Two photodiode signals are sent out from the sensor, one is the transmitted pump beam signal, and the other one is the reference signal by recording a part of the input beam before it enters the cavity. This latter one is also used to lock the time averaged input beam power to 0.14~mW. The current signals from these detectors are first converted to voltage signals by two transimpedance amplifiers with feedback resistors of 10 k$\Omega$ and bandwidths around 100~Hz. These voltage signals are sent to a differential amplifier, with a unit gain on the pump beam signal channel and a variable gain $g$ on the reference signal channel. The subtractor output is further amplified by a factor of 200, and recorded as the magnetometer signal.

Two coordinates are used to describe the spin dynamics following the notations in Ref.~\cite{cai2020}. One is the $xyz$ coordinate with the $z$ axis along the center of pump beam directions. The other one is $x'y'z'$ coordinate with the $z'$ axis along the bias field direction, and $y'$ axis perpendicular to the $zoz'$ plane. We denote the angle between the $z$ and $z'$ axes as $\psi_z$. For this single-beam scalar magnetometer, we fix the $y$ axis in the $zoz'$ plane as shown in the inset of Fig.~\ref{fig:scalarsetup}. The dynamics of the electron spin polarization $\boldsymbol{P}$ is described by the Bloch equation~\cite{shah09}:
\begin{equation}~\label{eq:bloch}
\frac{d\boldsymbol{P}}{dt}=\gamma\boldsymbol{P}\times(\boldsymbol{B}_0+\boldsymbol{B}_L)+\frac{1}{Q(P)}[R_{OP}(\boldsymbol{s}-\boldsymbol{P})-R_d\boldsymbol{P}],
\end{equation}
where $\gamma$ is the atomic gyromagnetic ratio, $Q(P)$ is the nuclear spin slowing down factor, $R_{OP}$ is the optical pumping rate, $\boldsymbol{s}$ is the photon spin of the light beam~\cite{appelt98}, $R_d$ is the atom depolarization rate in the absence of light, and $\boldsymbol{B}_L$ is the effective field induced by the light shift effect. $B_L$ is connected to $R_{OP}$ by the relation~\cite{appelt1999,savukov05}
\begin{equation}~\label{eq:bl}
R_{OP}s+i(2I+1)\gamma B_L=\frac{r_ecf\Phi s/Ar}{\Gamma-i\Delta\nu},
\end{equation}
with $I$ as the nuclear spin, $r_e$ as the classical electron radius, $c$ as the speed of light, $f$ as the oscillator strength, $\Phi$ as the light photon flux, $Ar$ as the beam area, $\Gamma$ as the atomic transition line width, and $\Delta\nu$ as the light detuning.

For a pump beam which is amplitude modulated at a frequency $\omega$, the optical pumping rate can be expressed as
\begin{equation}~\label{eq:Rop}
R_{OP}=a_0+\sum_{n=1}^\infty a_n\cos(n\omega t-\alpha_n),
\end{equation}
where $a_i$ is the corresponding coefficient of the Fourier expansion series of $R_{OP}$. With the pump beam duty cycle of 20\% in this experiment, we have $a_1/a_0\simeq1.9$. When the pump beam is on resonance with the Rb D1 transition and $\omega$ is close to the atomic Larmor precession frequency $\omega_L$, a substantial transverse atomic polarization can be built as~\cite{bell1961,cai2020}
\begin{equation}~\label{eq:pt}
P_{x'}+iP_{y'}=\frac{sr_na_1\sin\psi_z}{2[R+iQ(P)(\omega_L-\omega)]}e^{-i(\omega t-\alpha_1)},
\end{equation}
where $R=R_d+a_0$, $r_n$ is the abundance of the Rb isotope that is driven by the modulated pump beam. In most situations, $\omega\gg R$ and the ac part of longitudinal atomic polarization is negligible compared with $P_{x'}$~\cite{cai2020}. The static part of the longitudinal polarization is:
\begin{equation}~\label{eq:pz}
P_{z'}=sa_0\cos\psi_z/(R_d+a_0)
\end{equation}

The propagation equation of the pump beam intensity is:
\begin{equation}~\label{eq:dIz}
\frac{dI(z)}{dz}=-I(z)[1-\boldsymbol{s}\cdot{\boldsymbol{P}(z)}]n\sigma,
\end{equation}
where $\sigma$ is the photon absorption cross section of atoms, and $n$ is the atom density. Taking the time average, we have:
\begin{eqnarray}~\label{eq:Iz}
\frac{d\langle I(z)\rangle}{dz}=-\langle I(z)\rangle n\sigma\left[1-sP_{z'}(z)\cos\psi_z-\right.\nonumber\\
\left.\frac{sr_na_1(z)\sin\psi_z}{2a_0(z)}A({P}_{x',s}(\omega))\right],
\end{eqnarray}
where the operator $\langle\cdots\rangle$ denotes the time average, $A(P)$ is the amplitude of the oscillating parameter $P$, and $P_{x',s}$ is the part of $P_{x'}$ that is in synchronization with the first harmonic of the pump beam modulations in Eq.~\eqref{eq:Rop},
\begin{equation}~\label{eq:Px'}
P_{x',s}(\omega)=\frac{sa_1\cos(\omega t-\alpha_1)\sin\psi_z}{2}\frac{R}{R^2+Q^2(P)(\omega_L-\omega)^2}.
\end{equation}

The analytic solution of Eq.~\eqref{eq:Iz} is the principal value of the Lambert W function~\cite{kornackthesis}. A more convenient way to effectively express the transmitted beam intensity $I_t$ is:
\begin{equation}~\label{eq:It}
\langle I_t\rangle=c\langle I_0\rangle e^{-N_d}e^{N_d\left[s\overline{P}_{z'}\cos\psi_z+sr_n\sin\psi_zA(\overline{P}_{x',s}(\omega))\overline{a}_1/2\overline{a}_0\right]},
\end{equation}
where $I_0$ is the input beam intensity, the coefficient $c$ denotes the ratio of $I_t/I_0$ without atoms, and $c$ is determined by the optical patterns and beam power loss inside the cavity. The optical depth $N_d$ is equal to $n\sigma l$, with $l$ as the length of optical pathes inside the cell, and $\overline{x}$ is the spatial average of the parameter $x$ over the whole optical pathes.


\section{Magnetometer results and problems}
\begin{figure}[htb]
\includegraphics[width=3.0in]{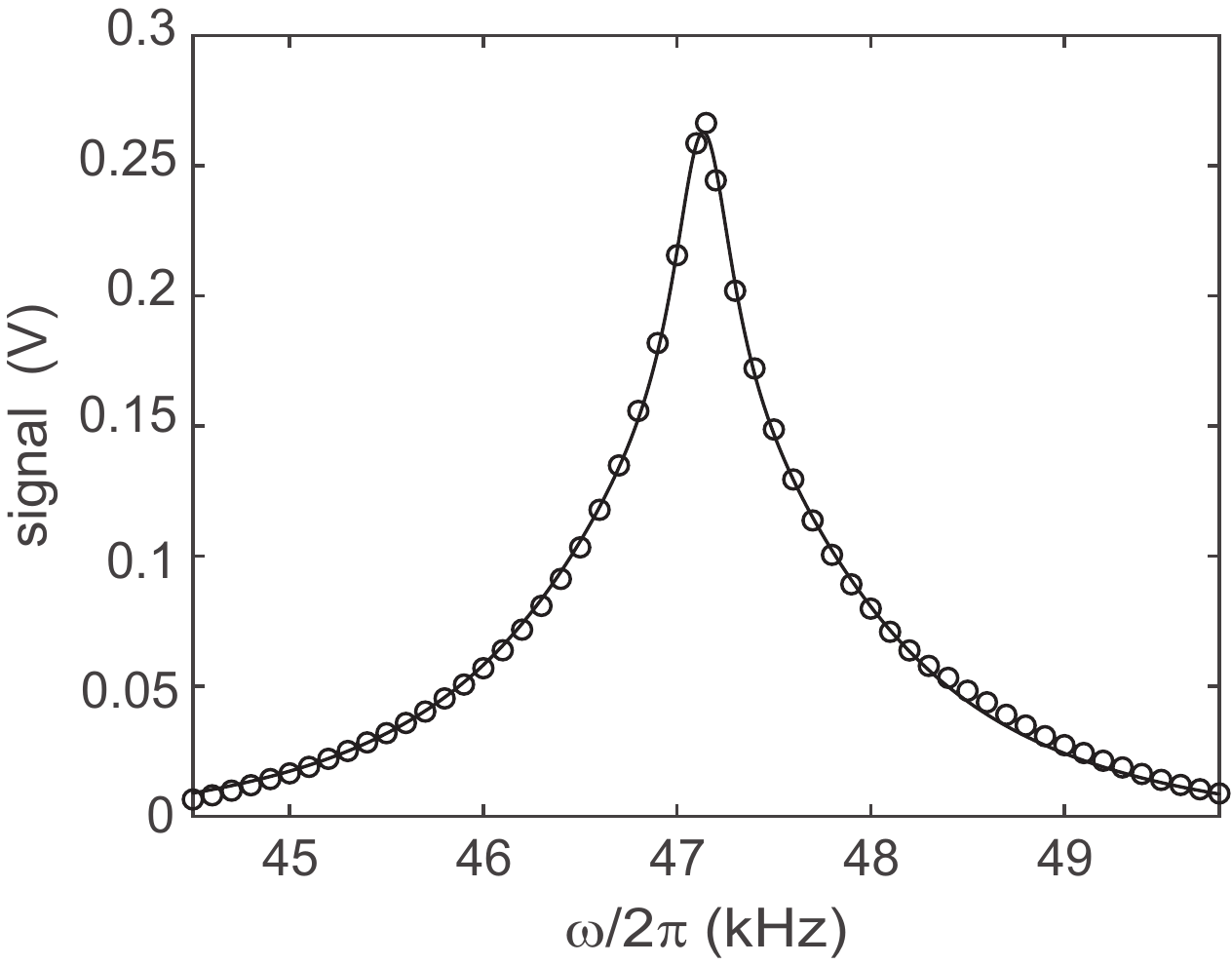}
\caption{\label{fig:result1A} Illustration of the magnetometer signal as a function of the pump beam amplitude modulation frequency while the pump beam is on resonance with the Rb D1 line and $\psi_z = 30^\circ$.  Here we focus on $^{85}$Rb atoms, as follows in the rest of the paper. The solid line is the fitting result using Eq.~\eqref{eq:fl}.}
\end{figure}

In this scalar magnetometer, the bias field amplitude is extracted from the atomic Larmor frequency, which is determined by finding the resonant response of atoms to $\omega$. Figure~\ref{fig:result1A} shows the experiment results when the pump beam is resonant with the Rb D1 line. Here, the recorded signal can be expressed as $V=G(\langle I_t\rangle-g\langle I_{ref}\rangle)$, where $I_{ref}$ is the reference beam intensity recorded by one of the photodiode detectors in the sensor, $g$ is the variable gain of the reference signal channel in the differential amplifier mentioned previously, and $G$ is the converting factor that connects the recorded beam intensity and the voltage output from the sensor electronics. The data shows symmetrical line shapes, which agrees with predictions from Eqs.~\eqref{eq:Px'} and ~\eqref{eq:It}. The exact line shape of the Herriott-cavity-assisted magnetometer signal is complex, especially due to the diffusion of atoms among the complicated optical patterns~\cite{xiao2006,sheng2013,lucivero2017}. It has been shown that~\cite{li2011} we can use a combination of two Lorentzian functions to describe the atomic polarization in the frequency domain.  Therefore, the magnetometer signal can be fitted by the following equation:
\begin{equation}~\label{eq:fl}
f(\omega)=h\exp\left[\sum_{i=1}^{2}\frac{c_i(\frac{\Gamma_i}{2})^2}{(\omega-\omega_0)^2+(\frac{\Gamma_i}{2})^2}\right]+b.
\end{equation}
Figure~\ref{fig:result1A} shows good fitting results using Eq.~\eqref{eq:fl}.

To characterize the magnetometer signal amplitude, we define a parameter $S$ as the difference between the magnetometer resonant signal ($\omega=\omega_L$) and the signal when $\omega$ is far from $\omega_L$. In the case that the light is resonant with the Rb D1 line, $S$ is determined from Eq.~\eqref{eq:It}:
\begin{eqnarray}~\label{eq:samp}
S&\propto&\langle I_t(\omega=\omega_L)\rangle-\langle I_t(\omega\rightarrow\infty)\rangle\nonumber\\
&\propto&\exp\left[-N_d+N_d\frac{s^2\overline{a_0}}{R_d+\overline{a_0}}\cos^2\psi_z\right]\times\nonumber\\
&&\left[\exp\left(N_d\frac{s^2r_n\overline{a_1}}{4\overline{a_0}}\frac{\overline{a_1}}{R_d+\overline{a_0}}\sin^2\psi_z\right)-1\right].
\end{eqnarray}

\begin{figure}[htb]
\includegraphics[width=3.0in]{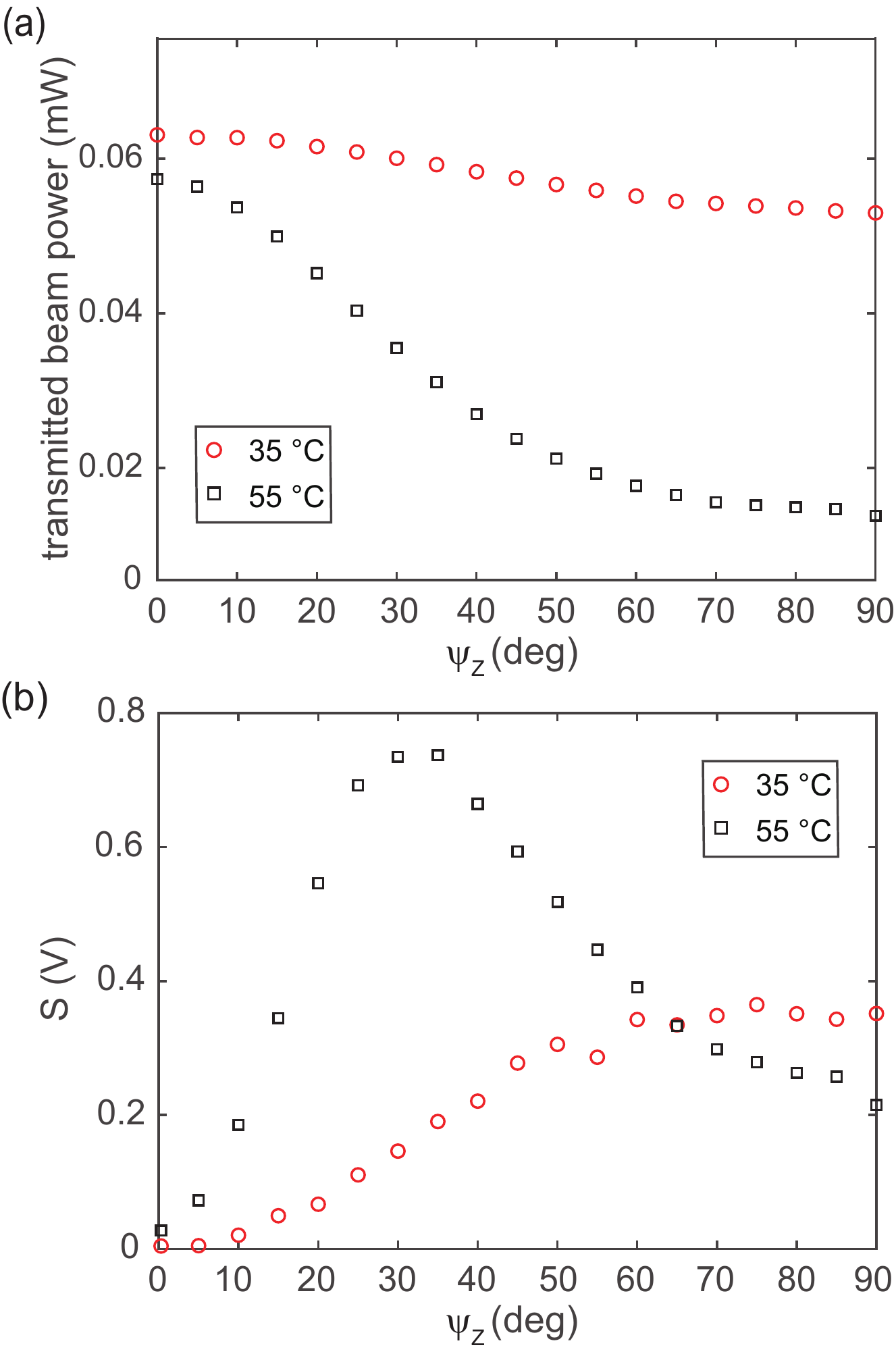}
\caption{\label{fig:result1B}(Color online) Plot~(a) and (b) shows the transmitted beam power and the $S$ value of the magnetometer response as a function of the $\psi_z$ at two different cell temperature.}
\end{figure}

When the cell temperature is low enough, the optical depth $N_d$ is negligible and the atomic polarization part in Eq.~\eqref{eq:Iz} can be neglected, so that the transmitted beam intensity is almost independent of the atom polarization. This is the situation of the experiment data at cell temperature of 35$^\circ$C in Fig.~\ref{fig:result1B}, where the transmitted beam power has weak angular dependence as shown in Fig.~\ref{fig:result1B}~(a). In this case, $S$ is determined by the last line in Eq.~\eqref{eq:samp}, which predicts that $S$ is maximum at $\psi_z=90^\circ$ and decreases with $\psi_z$. This is consistent with the experiment results at the same cell temperature in Fig.~\ref{fig:result1B}~(b). As the cell temperature and $N_d$ increases, the atomic polarization part starts to play an important role to determine the transmitted beam power. This case corresponds to the experiment data at cell temperature of 55$^\circ$C in Fig.~\ref{fig:result1B}.  Due to the $\cos^2\psi_z$ term in Eq.~\eqref{eq:samp}, the maximum amplitude of $S$ shifts from $\psi_z=90^\circ$ as shown in Fig.~\ref{fig:result1B}~(b). While the two experiment conditions lead to quite different experiment results when $\psi_z$ is large, the magnetometer signal amplitudes show similar rapid decay as $\psi_z$ approaches to zero. This is due to the last line in Eq.~\eqref{eq:samp}, which is proportional to $\sin^2\psi_z$ when $\psi_z$ is small. Therefore, the regions near $\psi_z=0^\circ$ are detection dead zones of current magnetometers.

\begin{figure}[htb]
\includegraphics[width=3in]{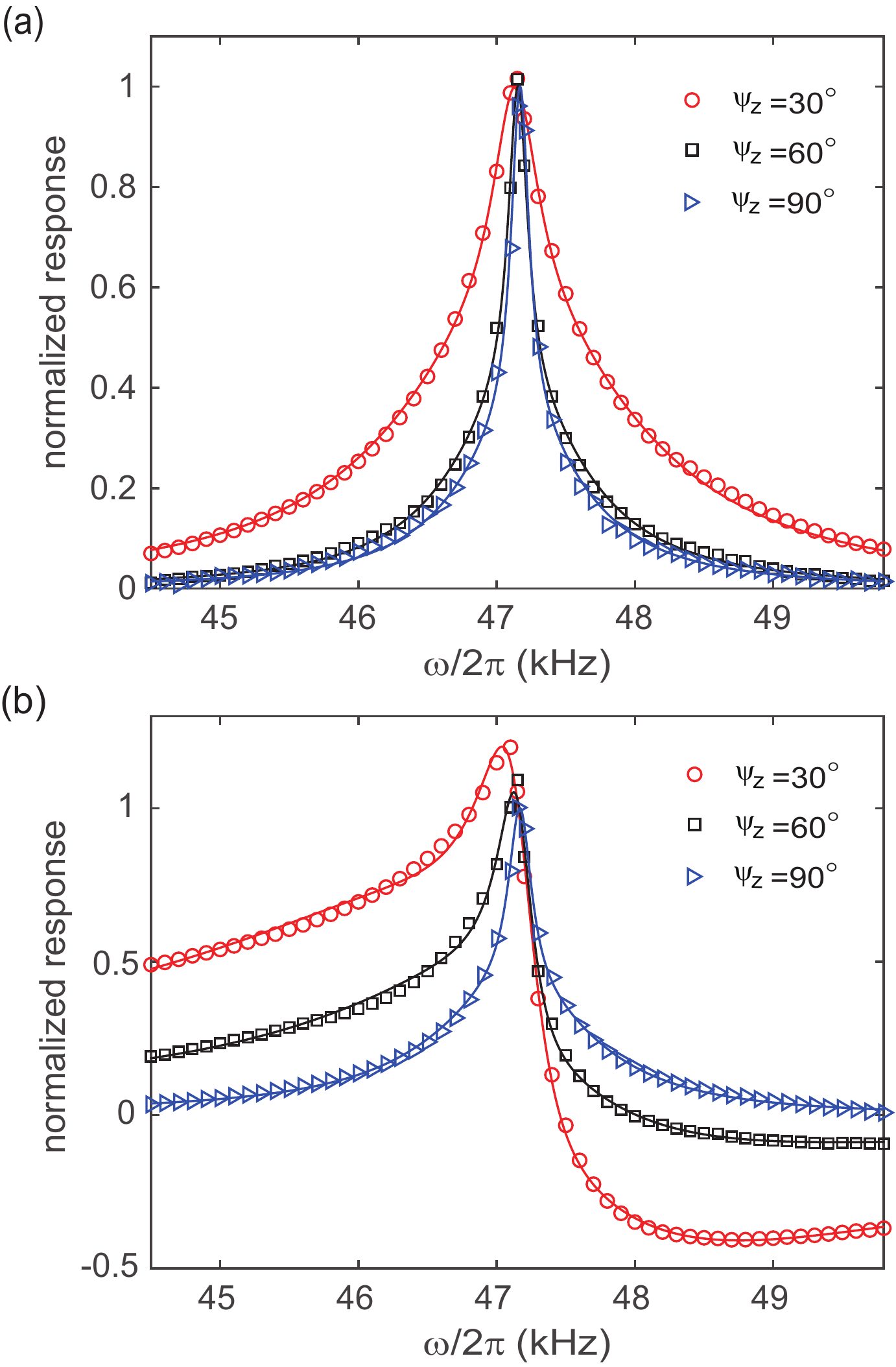}
\caption{\label{fig:result2}(Color online) (a) Normalized magnetometer responses at different $\psi_z$, when the pump beam is resonant with the Rb D1 line and the cell temperature is around 55 $^\circ$C. The solid lines are fitting results using Eq.~\eqref{eq:nw}. Plot~(b) shows the same results as plot~(a), except that the pump beam is 6 GHz blue detuned from resonance. The solid lines are fitting results using Eq.~\eqref{eq:nw2}.}
\end{figure}

To better compare the experiment results in different experiment conditions, we further define the normalized magnetometer response as the original magnetometer signals modified by subtracting the signal when $\omega$ is far off resonance, and then normalized by the parameter $S$. For the data in Fig.~\ref{fig:result1A}, its corresponding normalized magnetometer response can be described by:
\begin{eqnarray}~\label{eq:nw}
N(\omega)&=&\frac{f(\omega)-(h+b)}{S}\nonumber\\
&=&\frac{\exp\left[\sum_{i=1}^{2}\frac{c_i(\frac{\Gamma_i}{2})^2}{(\omega-\omega_0)^2+(\frac{\Gamma_i}{2})^2}\right]-1}{\exp(\sum_{i=1}^2 c_i)-1},
\end{eqnarray}
where the parameters in the equation are the same as Eq.~\eqref{eq:fl}. Figure~\ref{fig:result2}~(a) shows the normalized magnetometer responses with three different $\psi_z$ at a cell temperature around 55 $^\circ$C. While all the results have symmetric line shapes, the line width increases as $\psi_z$ decreases. This is due to fact that, as shown in Fig.~\ref{fig:result1B}~(a), the transmitted beam power increases when $\psi_z$ changes from 90$^\circ$ to $30^\circ$, which leads to a larger power broadening effect.

However, the situation becomes more complicated when the pump beam is off resonance. In this case, we need to consider the light shift effects. Similar to $R_{OP}$, the effective field from light shifts can be expanded as $\boldsymbol{B}_L=[B_{L0}+\sum_nB_{Ln}\cos(n\omega t-\alpha_n)]\hat{\boldsymbol{z}}$. The first harmonic part of $\boldsymbol{B}_L$ can excite a part of $P_{z'}$ to the $x'y'$ plane, so that there is an additional polarization term $P'_{x'}$ along the $x'$ axis besides $P_{x'}$ in Eq.~\eqref{eq:pt}. By treating this effect as a perturbation, we can express $P'_{x'}$  as~\cite{seltzerthesis}:
\begin{eqnarray}~\label{eq:px2}
P_{x'}'=\frac{(\omega_L-\omega)\gamma B_{L1}\cos(\omega t-\alpha_1)\sin\psi_zP_{z'}}{2[R^2/Q^2(P)+(\gamma B_{L1}\sin\psi_z/2)^2+(\omega_L-\omega)^2]}.
\end{eqnarray}
The transmitted pump beam intensity is now determined by $A(\overline{P}_{x',s}+\overline{P'}_{x'})$, instead of $A(\overline{P}_{x',s})$, in Eq.~\eqref{eq:It}. Therefore, the dispersion relation between $P_{x'}'$ and $\omega$ introduces asymmetric magnetometer response results. Similar to Eq.~\ref{eq:fl}, a combination of two Lorentzian functions and two dispersion functions are used to describe the atomic polarization behaviors. The function to fit the magnetometer signal with detuned light is:
\begin{eqnarray}~\label{eq:fld}
F(\omega)=h\exp\left[\sum_{i=1}^{2}\frac{c_i(\frac{\Gamma_i}{2})^2+d_i(\omega-\omega_0)}{(\omega-\omega_0)^2+(\frac{\Gamma_i}{2})^2}\right]+b,
\end{eqnarray}
and the normalized response is described by:
\begin{eqnarray}~\label{eq:nw2}
N'(\omega)&=&\frac{F(\omega)-(h+b)}{S}\nonumber\\
&=&\frac{\exp\left[\sum_{i=1}^{2}\frac{c_i(\frac{\Gamma_i}{2})^2+d_i(\omega-\omega_0)}{(\omega-\omega_0)^2+(\frac{\Gamma_i}{2})^2}\right]-1}{\exp(\sum_{i=1}^{2}c_i)-1}.
\end{eqnarray}

Figure~\ref{fig:result2}~(b) shows the normalized magnetometer results, when the pump beam is 6 GHz blue detuned from the D1 line. It confirms the appearance of asymmetrical line shapes as $\psi_z$ deviates from 90$^\circ$. On the other hand, the data also shows that the asymmetry of the line shapes increases as $\psi_z$ decreases. This is because that the degree of this asymmetry is determined by the ratio $A(\overline{P'}_{x'})/A(\overline{P}_{x',s})$, which is proportional to $\cos\psi_z$. In order to quantify the degree of the line shape asymmetry, we introduce a new parameter $D$, which is the ratio of the asymmetric part over the symmetric part in the fitting function:
\begin{eqnarray}~\label{eq:R}
D(\delta f)=\frac{\sum\limits_{i=1}^{2}\int_{\omega_L-2\pi\delta f}^{\omega_L+2\pi\delta f} \lvert\frac{d_i(\omega-\omega_0)}{(\omega-\omega_0)^2+(\frac{\Gamma_i}{2})^2}\rvert d\omega}{\sum\limits_{i=1}^{2}\int_{\omega_L-2\pi\delta f}^{\omega_L+2\pi\delta f}\lvert \frac{c_i(\frac{\Gamma_i}{2})^2}{(\omega-\omega_0)^2+(\frac{\Gamma_i}{2})^2}\rvert d\omega}.
\end{eqnarray}
The parameters in the equation above are the same as Eq.~\eqref{eq:fld}. Using the fitting results, we have $D(3000)=2.38$ for the data at $\psi_z=30^\circ$ in Fig.~\ref{fig:result2}~(b).

\section{Solutions and discussions}
The field-orientation dependent asymmetric magnetometer response, induced by ac light shifts, presents a new source of heading error in Bell-Bloom magnetometry. From Eqs.~\eqref{eq:Px'} and ~\eqref{eq:px2}, we can get that the ac light shift effect on the transmitted beam intensity is determined by $sB_{L1}$, the product of the photon spin and the effective field from the light shift. It is confirmed by the experiment that the asymmetry of the magnetometer response can be reversed by flipping the pump beam polarization or detuning. Therefore, to solve this heading error problem without consuming additional power, a simple way is to reverse the pump beam polarization inside the atomic cell.

If the optical length of the pump beam inside the cell with two opposite circular polarizations are same, the effect of $\boldsymbol{s}\cdot\boldsymbol{P}'_{x'}$ in Eq.~\eqref{eq:dIz} can be cancelled out. In the same way, the static light shift can also be spatially averaged out. Therefore, the whole light shift effects in this magnetometry can be eliminated using such a method. In practice, to maintain the magnetometer signal, it is also important to reduce the possibility that atoms polarized by one pump beam polarization diffuse into another region pumped by light with an opposite polarization. Considering all the requirements, the most suitable way to implement this scheme in the experiment is to attach a half-wave plate in the middle of the cavity.

In the experiment, we use a customized true zero-order half-wave plate, which is attached to a piece of K9 glass so that the total thickness of the wave plate is 0.6~mm. Due to the heated environment around the cell, we need to first check the temperature effect on the performance of these half-wave plates. For a transverse wave propagating along the $z$ direction, the oscillation amplitude of its electric field can be expressed as $A_{x,y}=E_{x,y}\cos(\omega t+\delta_{x,y})$. The circular polarization part in this wave can be characterized by an ellipticity parameter $\eta$~\cite{born1999}:
\begin{equation}~\label{eq:eta}
\sin(2\eta)=\frac{2{E}_{x}{E}_{y}\sin(\delta_x-\delta_y)}{|{E}_{x}|^2+|{E}_{y}|^2}.
\end{equation}
An ideally circularly polarized wave has $\eta=45^\circ$, and a linearly polarized wave has $\eta=0^\circ$.

\begin{figure}[htb]
\includegraphics[width=3in]{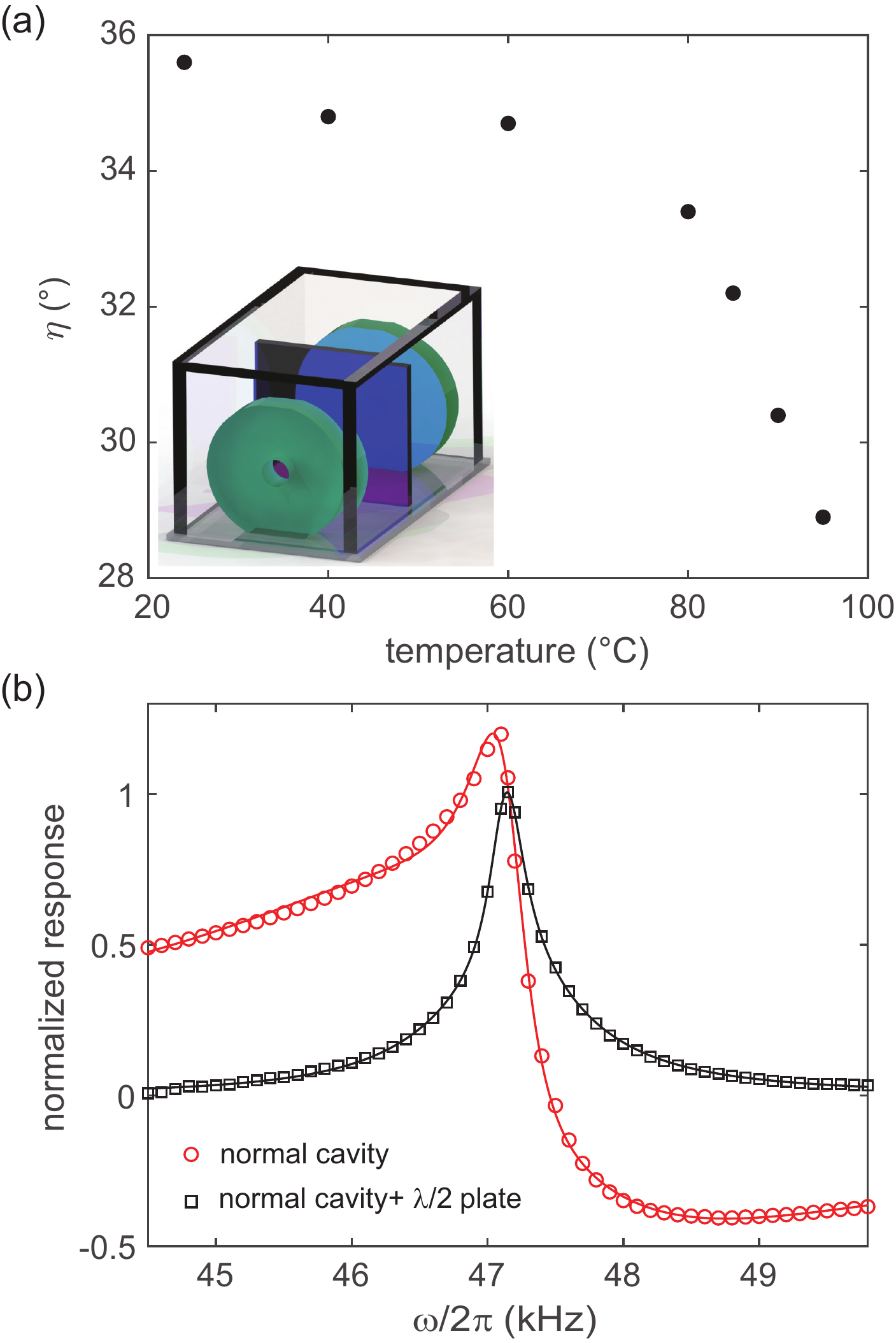}
\caption{\label{fig:result3}(Color online) Plot~(a) shows the ellipticity parameter $\eta$ of the output beam as a function of the cell temperature, with the cavity configuration shown in the inset. Plot(b) compares the magnetometer results using two different cavity configurations. The lines are fitting curves  using Eq.~\eqref{eq:nw2}.}
\end{figure}

We attach one of the half-wave plates in the middle of the cavity as shown in the inset of Fig.~\ref{fig:result3}~(a), use a far-off-resonance light with $\eta$ close to 45$^\circ$ as the input beam, and measure $\eta$ of the output beam from the cavity at different cell temperature. The experiment data in the same plot shows that the output beam has a weak elliptical polarization at low temperature, due to 21 reflections on the cavity mirrors and 22 passes through the wave plate. $\eta$ of the output beam changes slightly when the temperature is below 80 $^\circ$C, and the circular polarization of the output beam rapidly degrades when the cell temperature is above 85 $^\circ$C. Therefore, the customized half-wave plates are feasible for applications when the cell temperature is below 80 $^\circ$C, which covers the sensor operation temperature in this paper.

Using this modified cavity with a half-wave plate in the middle, we retake the magnetometer results at $\psi_z=30^\circ$. The new experiment data in Fig.~\ref{fig:result3}~(b) shows $D'(3000)=0.09$. Compared with the results using the conventional cavity in the previous section, we have eliminated most of the aforementioned light shift effect, and only $3.7\%$ of the original effect is left. This residual effect can be further reduced by tuning the position of the half-wave plate inside the cavity.

\begin{figure}[htp]
\includegraphics[width=3in]{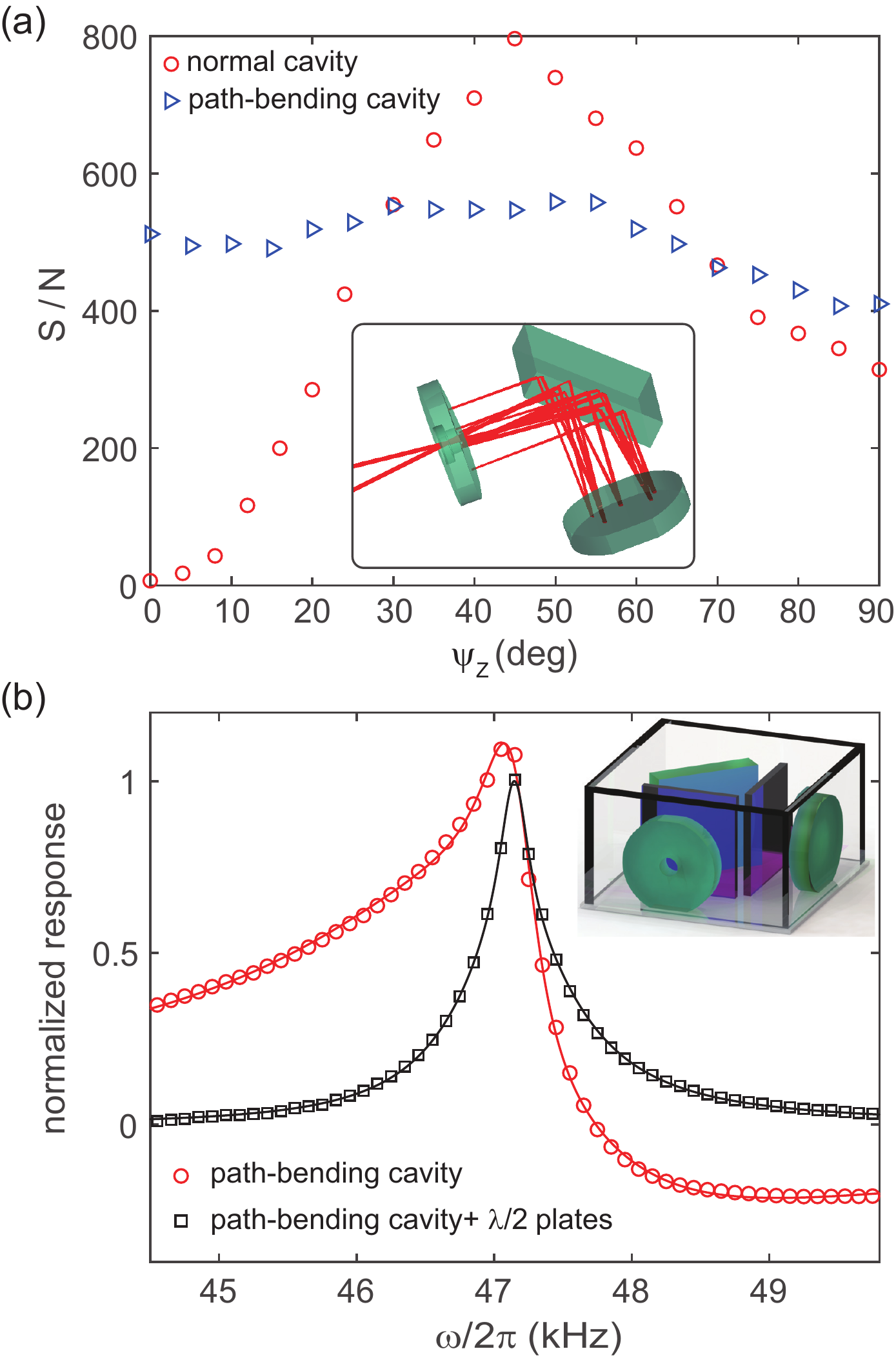}
\caption{\label{fig:result4}(Color online) Plot~(a) shows the magnetometer S/N as a function of the $\psi_z$  using two cavity configurations with the pump beam resonant with the Rb D1 line, $\omega=\omega_L$, and cell temperature around 55 $^\circ$C. The inset shows an illustration of the path-bending cavity. Plot~(b) shows the normalized response of the magnetometers taken at $\psi_z=45^\circ$, with the pump beam 6 GHz blue detuned from the Rb D1 line. The lines are fitting curves using Eq.~\eqref{eq:nw2}, and the inset illustrates a path-bending cavity vapor cell with half-wave plates.}
\end{figure}

We use the parameter S/N as the magnetometer signal-to-noise ratio with a 1 Hz bandwidth. Figure~\ref{fig:result4}~(a) demonstrates that there are two orders of magnitude difference for the maximum and minimum of S/N in the magnetometer using a conventional cavity. To lift the detection dead zones in the sensor, we add a reflection mirror to the conventional cavity, and arrange the cavity mirrors as shown in the inset of Fig.~\ref{fig:result4}~(a). In this new configuration, the optical paths are bent by 90$^\circ$ in the middle. This path-bending cavity can be treated as a combination of two separated normal multipass cavities. They can complement each other because their dead zones are orthogonal.

While it helps to eliminate the detection dead zones, the introduction of an extra reflection mirror also doubles the interactions between the mirror surfaces and the beam, which can strongly influence the quality of the beam circular polarization. In practice, we modify the cavity mirror parameters by reducing the number of reflections inside the cavity to 13 times, and increasing the length of each optical path inside the cavity to 26.3~mm. In this way, the magnetometer signal amplitude is kept almost unchanged, and the polarization of the beam inside the cavity is also maintained in a good quality. As shown in Fig.~\ref{fig:result4}~(a), S/N of the magnetometer using this path-bending cavity fluctuates within 20\% over the whole range of $\psi_z$. This is a significant improvement compared with the magnetometer using a conventional cavity, and the remaining orientation dependence of the S/N is mainly determined by the cavity structure.

It is straightforward to combine the two methods developed above for simultaneous elimination of the dead zones and light shifts. As shown in the inset Fig.~\ref{fig:result4}~(b), this is achieved by placing two half-wave plates in the middle of the optical paths between the reflection mirror and cavity mirrors. Its function is confirmed by the experiment results in the Fig.~\ref{fig:result4}~(b), which shows similar success on suppressing the light-shift effect using this new configuration as the one in Fig.~\ref{fig:result3}~(b).

\section{Conclusion}
In summary, we have demonstrated an atomic orientation based scalar magnetometer using a single amplitude-modulated beam and a multipass cell. We suppress the light-shift effect by more than one order of magnitude, using a half-wave plate in the middle of the cavity to reverse the beam polarization inside the cavity. We also remove the detection dead zones, and limit the change of the magnetometer signal-to-noise ratio within 20\% over the whole angular range, by modifying the cavity to a three-mirror configuration. Combining these two techniques, we reached a light-shift-free and dead-zone-free scalar magnetometer. The methods developed in this paper are novel in optically pumped magnetometry, robust to use, and easy to be extended to other atomic devices.

One important application is in the aforementioned atomic magnetometry using a single elliptically polarized light~\cite{shah09}, where this beam has to be kept off resonance for both optical pumping and Faraday rotation detections. The inherent light shift problem limits the application of this magnetometry in precise scalar field measurements. The polarization-reversing cavity developed in this paper can solve this problem by eliminating the first order light shift effect. Calculations based on Jones calculus have been performed to confirm the feasibility of this scheme. If the proposed scheme can be successfully implemented, this modified magnetometer would be a promising compact single-beam version to replace the widely used two-beam double-resonance scalar magnetometers~\cite{smullin2009}.

\section*{Acknowledgements}
This work was partially carried out at the USTC Center for Micro and Nanoscale Research and Fabrication. This work was supported by National Natural Science Foundation of China (Grant No. 11974329), and Scientific Instrument and Equipment Development Projects, CAS (NO. YJKYYQ20200043).

\end{document}